\documentclass[aps,prb,reprint,a4paper,floatfix,groupedaddress,amsmath,amssymb,amsfonts]{revtex4-1}
\pdfoutput=1
\usepackage{newtxtext,newtxmath}
\usepackage[utf8]{inputenx} 
\usepackage{graphicx}
\usepackage{xspace}
\usepackage[dvipsnames]{xcolor}
\usepackage{soul} % provides highlighting
\usepackage[pdftex]{hyperref}
\hypersetup{
	pdftitle={Fine structure of excitons in InAs/GaAs(110) quantum dots},
	colorlinks,
	citecolor=blue
	}
\usepackage[rightcaption]{sidecap}
\usepackage{color,soul} % provides highlighting
\usepackage[
,textwidth=17.5cm
,textheight=23.5cm
,verbose
,dvips
]{geometry}

\usepackage[colorinlistoftodos,prependcaption,textsize=footnotesize,textwidth=1.5cm]{todonotes}

\begin{document}
\title{Fine structure of excitons in InAs quantum dots on GaAs(110) planar layers
and nanowire facets}

\author{Pierre Corfdir}
\email{corfdir@pdi-berlin.de}
\author{Ryan B. Lewis}
\author{Lutz Geelhaar}
\author{Oliver Brandt}

\affiliation{Paul-Drude-Institut für Festkörperelektronik,
Hausvogteiplatz 5--7, 10117 Berlin, Germany}

%Sample used for this study: m3-1782 (planar sample) - m3-1701 (nanowire sample)

\begin{abstract}

We investigate the optical properties of InAs quantum dots grown by molecular beam epitaxy on GaAs(110) using Bi as a surfactant. The quantum dots are synthesized on planar GaAs(110) substrates as well as on the \{110\} sidewall facets of GaAs nanowires. At 10~K, neutral excitons confined in these quantum dots give rise to photoluminescence lines between 1.1 and 1.4~eV. Magneto-photoluminescence spectroscopy reveals that for small quantum dots emitting between 1.3 and 1.4~eV, the electron-hole coherence length in and perpendicular to the (110) plane is on the order of 5 and 2~nm, respectively. The quantum dot photoluminescence is linearly polarized, and both binding and antibinding biexcitons are observed, two findings that we associate with the strain in the (110) plane This strain leads to piezoelectric fields and to a strong mixing between heavy and light hole states, and offers the possibility to tune the degree of linear polarization of the exciton photoluminescence as well as the sign of the binding energy of biexcitons. 

\end{abstract}

\maketitle

\section{Introduction}

The discrete energy levels of semiconductor quantum dots (QDs) have made possible new classes of solid-state-based quantum devices such as single photon and entangled photon pair emitters\cite{Michler2000,Akopian2006,Stevenson2006} as well as all-optical logic gates.\cite{Li2003} A prerequisite for these applications is a high degree of control not only on the dimensions of the QDs, but also on their symmetry. For instance, entangled photon pairs can be produced in highly symmetrical QDs for which the anisotropic exchange splitting of the bright states of the exciton is smaller than the radiative linewidth.\cite{Benson2000,Schliwa2009} This goal is difficult to achieve using QDs on GaAs(001), where the different adatom mobilities along the [110] and $[1\bar{1}0]$ directions lead to elongated QDs whose symmetry is thus reduced from $D_{2d}$ to $C_{2v}$, inducing a finite fine structure splitting of the bright exciton states.\cite{Bayer2002} In contrast, group-III-arsenide-based QDs grown on GaAs(111) substrates patterned with pyramidal recesses exhibit a $C_{3v}$ symmetry resulting in a fine structure splitting of zero.\cite{Juska2013,Juska2014}

The above example illustrates that the fundamental electronic and optical properties of QDs as well as their specific technological applications are intimately related to their symmetry. It is thus of considerable interest to investigate QD systems with symmetries that differ from $C_{2v}$ and $C_{3v}$. Accordingly, the growth of GaAs-based QDs on high-index surfaces such as (113) and (115) has been explored.\cite{Mano2008,Lei2016} More recently, we reported that a Bi-surfactant-induced morphological instability can enable the growth of InAs 3D islands on GaAs(110), a surface on which 3D islands do not normally form.\cite{Lewis2017a,Lewis2017b} Due to the inequivalent $[1\bar{1}0]$ and [001] in-plane directions, these InAs(110) QDs are of $C_s$\cite{Singh2013} symmetry, and their optical properties are expected to differ significantly from that of $C_{2v}$ and $C_{3v}$ QDs. In particular, it has been predicted that the strong in-plane piezoelectric fields in (In,Ga)As(110) QDs modify their electronic structure, and that the light emission associated with the ground state bright exciton is linearly polarized.\citep{Singh2013}

Here, we use photoluminescence (PL) spectroscopy to investigate the optical properties of InAs QDs grown by molecular beam epitaxy on GaAs(110) substrates as well as on the \{110\} sidewall facets of GaAs nanowires. For this surface, the formation of QDs is usually inhibited since two-dimensional Frank-van der Merwe growth prevails regardless of the thickness of the strained InAs film, and strain relaxation occurs by the formation of misfit dislocations.\cite{Belk1997} To induce the Stranski-Krastanov growth of three-dimensional islands, we have used Bi as a surfactant.\cite{Lewis2017a,Lewis2017b}  From magneto-PL experiments, we show that for small InAs QDs emitting between 1.3 and 1.4~eV, the electron-hole coherence length of the exciton in these QDs is on the order of 5 and 2~nm in the in- and out-of-plane directions, respectively. As a result of the low $C_s$ symmetry of InAs(110) QDs, strain in the (110) plane leads to a strong mixing between heavy and light holes, as well as to strong piezoelectric fields. While the former is promising for the fabrication of single photon emitters with a high degree of linear polarization, the latter is of interest to obtain QDs with zero biexciton binding energy for the emission of entangled photon pairs.

\begin{SCfigure*}
\includegraphics[scale=1]{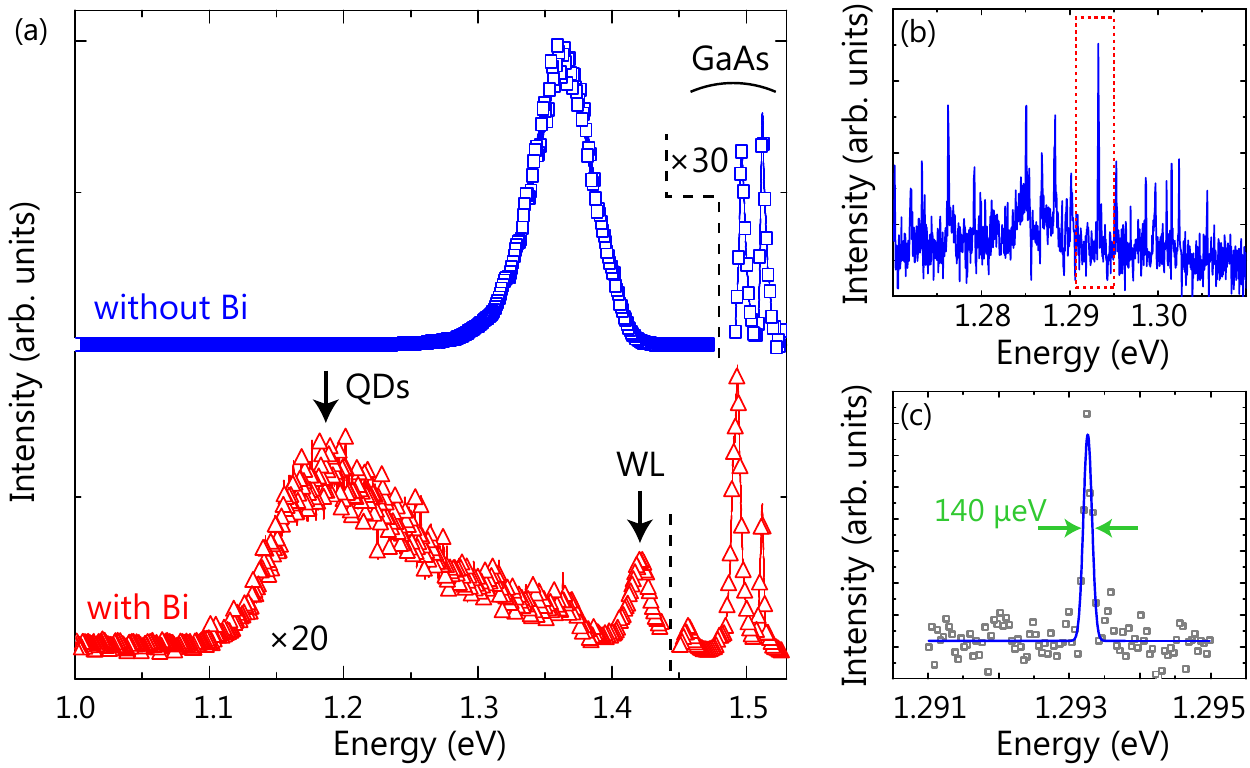}
\caption{(a) PL spectra acquired at 10~K from the planar InAs/GaAs(110) sample grown with (triangles) and without (squares) the Bi surfactant. The spectra have been normalized and shifted vertically for clarity. The transitions related to the GaAs substrate, the WL, and the InAs QDs are indicated on the figure. (b) PL spectrum at 4.2~K of the sample grown with a Bi surfactant. (c) Enlarged view of the transition highlighted by a red rectangle in (b). The solid line is a Gaussian fit yielding a full width at half maximum of 140~\textmu eV.}
\label{fig:Figure1}
\end{SCfigure*} 

\section{Experimental details}

InAs QDs were synthesized by molecular beam epitaxy on planar GaAs(110) substrates and on the $\{1\bar{1}0\}$ sidewalls of GaAs nanowires grown on Si(111) substrates. For the planar samples, about 1.5 monolayers of InAs were deposited on a 150~nm thick GaAs buffer layer and subsequently capped with 50~nm of GaAs. For the nanowire samples, an equivalent amount of InAs was deposited onto the sidewalls of GaAs nanowire cores of about 60~nm diameter, 7~\textmu m length, and a density of 0.3~\textmu m$^{-2}$. The nanowires were then clad by a GaAs/AlAs/GaAs multishell structure with respective thicknesses of 5/10/5~nm. For both planar and nanowire samples, a Bi flux (with a beam equivalent pressure of $2\times 10^{-6}$~mbar) was present during the InAs deposition, which was carried out at 420\,°C. The presence of the Bi surfactant modifies the surface energies, inducing the formation of three-dimensional InAs islands by a process resembling Stranski-Krastanov growth. Comparison samples were also grown without the Bi flux, and in these samples QDs did not form. For the planar sample, the rotation of the substrate was stopped during InAs growth, leading to a gradient in the density of the QDs over the wafer. Further details of the growth process of the planar and nanowire samples can be found in Refs.~\onlinecite{Lewis2017a,Lewis2017b}, respectively.

All PL experiments have been carried using a Ti:Sapphire laser emitting at 790~nm as the excitation source. For measurements at 10~K, the samples were mounted on the coldfinger of a continuous-flow He cryostat, and the laser beam was focused using microscope objectives with numerical apertures of 0.25, 0.55 or 0.7. The PL was dispersed with a monochromator equipped with a 900~lines/mm grating and detected with a liquid N$_2$-cooled (In,Ga)As array or a charge-coupled device. The analysis of the polarization of the PL signal was performed using a polarizer followed by a half-waveplate. For measurements at 4.2~K, the samples were kept at liquid He temperature in a confocal setup, and the laser beam was focused using a microscope objective with a numerical aperture of 0.82. The PL signal was collected using the same objective and coupled to a single mode fiber, whose core acted as a confocal hole (the core diameter of the fiber was 4.4~\textmu m). The signal was dispersed using a 900~lines/mm grating and detected with a liquid N$_2$-cooled charge-coupled device. With this setup, magnetic fields $B$ with a strength between 0 and 8~T could be applied in Faraday configuration, i.\,e., $B||[110]$ and $B||[111]$ for the planar and the nanowire samples, respectively.

\section{Results and discussion}

Figure~\ref{fig:Figure1}(a) shows a PL spectrum taken at 10~K from the sample with InAs QDs on planar GaAs(110). A spectrum of a sample grown without Bi surfactant, while keeping other growth conditions the same is also displayed. Both samples exhibit emission lines centered at 1.513, 1.495 and 1.459~eV, originating from the GaAs bound exciton and the GaAs band-to-carbon transition along with its phonon replica, respectively. For both samples, the intensities of these lines are almost independent of the position of the excitation spot. For the sample grown without the Bi surfactant, a strong PL band at 1.362~eV is observed, which we attribute to charge carrier recombination in the quantum well formed by the two-dimensional InAs layer. In contrast, for the sample grown with Bi, two bands are observed on the lower energy side of GaAs-related PL lines. While the band at 1.422~eV exhibits a full width at half maximum of 17~meV, the band with a peak energy of 1.19~eV is much broader and exhibits an asymmetric lineshape. Following the result in Ref.~\onlinecite{Lewis2017a}, we attribute the bands at 1.422 and 1.19~eV to carrier recombination in the InAs wetting layer (WL) and QDs, respectively. As shown in Fig.~\ref{fig:Figure1}(b), the emission band related to the QDs consists of tens of narrow lines, each associated with an individual QD. These lines exhibit a full width at half maximum as narrow as 140~\textmu eV [see Fig.~\ref{fig:Figure1}(c)], corresponding to the spectral resolution of the setup. As the substrate rotation was stopped during the InAs deposition, the spatial density of QDs and the corresponding spectral density of lines at 1.3~eV vary over the sample. In the following, all experiments have been carried out on a region of the sample where the density of QDs emitting between 1.3 and 1.4~eV is sufficiently low to facilitate single QD spectroscopy.

\begin{figure}
\includegraphics[scale=1]{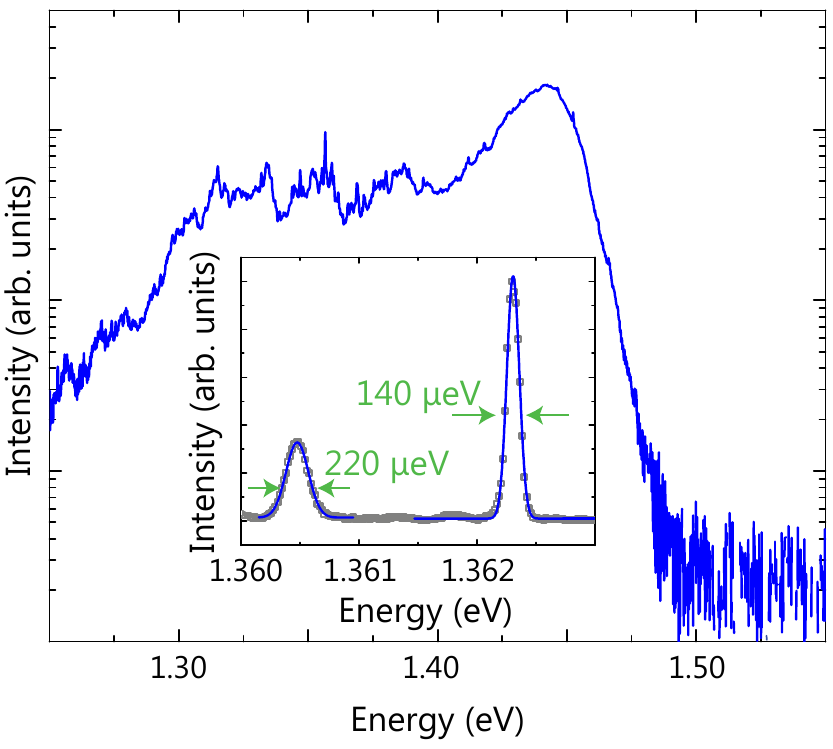}
\caption{PL spectrum obtained at 10~K from a few core-multishell nanowires grown with Bi surfactant. The inset shows an expanded view of two individual transitions. The full widths at half maxima indicated in the inset have been obtained by Gaussian fits (blue lines).}
\label{fig:FigureNW}
\end{figure}

Figure~\ref{fig:FigureNW} shows a PL spectrum taken at 10~K on as-grown core-multishell nanowires. We estimate that about 5 nanowires are probed simultaneously in this experiment. The spectrum consists of a broad band centered at 1.44~eV with tens of narrow transitions on its lower energy side. As shown in the enlarged spectrum in inset, the linewidth of the latter transitions ranges typically from 140~\textmu eV (resolution limit of our setup) to a few hundreds of \textmu eV. Polytypism in InAs shells was previously shown to cause photoluminescence lines broader than thoses observed in Fig.~\ref{fig:FigureNW}.\cite{Corfdir2016b} Since these transitions were only observed for samples grown using the Bi surfactant,\cite{Lewis2017b} they are associated with Bi induced InAs islands. In analogy to the planar case, we attribute them to emission from single QDs that form on the sidewalls of the nanowires due to the presence of the Bi surfactant, while we attribute the band at 1.44~eV to the InAs WL. The increased width of these lines with respect to the resolution limit is presumably resulting from spectral diffusion due to fast electrostatic fluctuations in the vicinity of the QDs.\cite{Fontana2014,Holmes2015,Corfdir2016} In contrast to the planar sample, no emission lines related to GaAs could be observed, indicating that the capture by the QDs of carriers photoexcited in the GaAs core is highly efficient. This result is a consequence of the core-shell geometry of our nanowires and has been reported previously for nanowires with (In,Ga)As shell quantum wells.\cite{Corfdir2016b}

\begin{SCfigure*}
\includegraphics[scale=1]{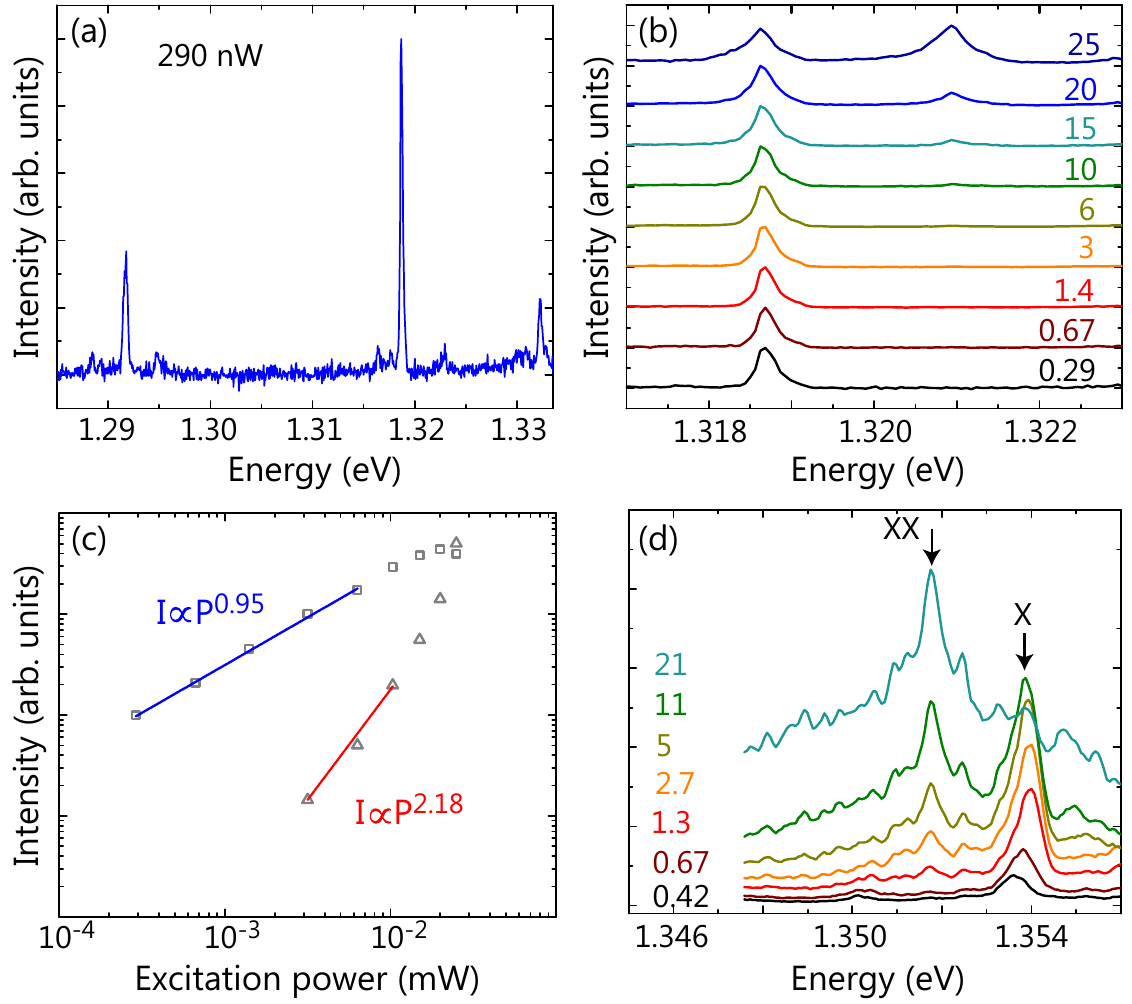}
\caption{(a) PL spectrum of a few InAs QDs on planar GaAs(110) taken at 4.2~K with an excitation power of 290~nW. (b) PL spectra from the QD emitting at 1.3187~eV in (a) for different excitation powers in \textmu W as indicated in the figure. The spectra have been normalized and shifted vertically for clarity. (c) Intensity of the transitions at 1.3187 and 1.3210~eV (squares and triangles, respectively) as a function of excitation power. The blue and red lines show fits yielding the slopes indicated in the figure. (d) PL spectra of a QD in a nanowire for different excitation powers. The excitation power in \textmu W is specified on the left of each spectrum. For excitation powers up to 2.7~\textmu W, the intensity of the lines labeled X and XX increases linearly and quadratically with increasing excitation power, respectively.}
\label{fig:Figure2}
\end{SCfigure*}

Figure~\ref{fig:Figure2}(a) shows a confocal PL spectrum taken on the planar QD sample at 4.2~K with an excitation power of 290~nW. Thanks to the increased spatial resolution, only a few QDs are excited, and individual transitions are well resolved. Figure~\ref{fig:Figure2}(b) presents a series of spectra recorded with different excitation powers of the line at 1.3187~eV in Fig.~\ref{fig:Figure2}(a). At high excitation powers, an additional line at 1.3210~eV appears in the spectra. Figure~\ref{fig:Figure2}(c) shows the dependence of the emission intensity of the lines at 1.3187 and 1.3210~eV on excitation power. Below a power of 10~\textmu W, the intensity of the line at 1.3187~eV increases linearly with increasing excitation power, demonstrating that it arises from the recombination of a neutral exciton. For higher powers, the intensity of the neutral exciton transition saturates and eventually decreases, a behavior already reported for other QD systems.\cite{Corfdir2016} In contrast, the intensity of the line at 1.3210~eV increases almost quadratically with increasing excitation power. This line is therefore related to an antibinding biexciton with a binding energy of $-2.3$~meV.

For our InAs QDs on planar GaAs(110), the biexciton transition energy was systematically 2--3~meV larger than that of the neutral exciton. To verify whether this observation is a general result for small InAs(110) QDs, we have performed PL experiments with varying excitation power also for the InAs QDs on the $\{ 1\bar{1}0 \}$ sidewalls of GaAs nanowires. Figure~\ref{fig:Figure2}(d) shows a typical spectrum of a single QD. With the excitation power increasing from 0.42 to 2.7~\textmu W, the intensity of the transition at 1.3536~eV increases linearly, indicating that it arises from a neutral exciton. The intensity of this line saturates at higher powers. In contrast to the planar case [Fig.~\ref{fig:Figure2}(b)], a biexciton transition appears for higher excitation powers at the lower energy side of the neutral exciton, i.\,e., the biexciton in Fig.~\ref{fig:Figure2}(d) has a binding energy of $+2.1$~meV. For InAs QDs on GaAs(001), the biexciton binding energy decreases with decreasing QD size.\citep{Rodt2003} However, this relation seems not to apply here, since the QD with binding biexciton in Fig.~\ref{fig:Figure2}(d) emits at an energy higher than the QD with an antibinding biexciton in Fig.~\ref{fig:Figure2}(b). We note that binding biexcitons were also reported for small InAs QDs deposited either on AlAs(110) using the cleaved edge overgrowth technique\cite{Uccelli2008} or on the $\{ 110 \}$ facets of GaAs/AlAs core-shell nanowires.\cite{Uccelli2010} 
 
To clarify the origin of the different character of the biexciton state, the actual shape and dimensions of the QDs need to be elucidated.\cite{KantaPatra2017} Atomic force micrographs of uncapped QDs are of limited relevance for this purpose, as significant in- and out-of-plane In segregation may occur during the QD overgrowth by GaAs.\cite{Giddings2011} The three-dimensional shape of a single embedded QD can be reconstructed by atom probe\cite{Giddings2011} or electron tomography,\cite{Verheijen2007,Niehle2015} but it is difficult to achieve results of statistical significance with these techniques. Magneto-PL is an alternative technique to obtain an estimate of the size of embedded InAs QDs or, rather, the spatial extent of the confining potential. In the presence of an external magnetic field $B$ and neglecting exchange splittings, the energy $E_\text{X}$ of a neutral exciton in a QD is given by:\cite{Bayer2002}

%=========================================
\begin{equation}\label{eq:shift}
E_\text{X} = E_\text{X}^0 \pm \frac{1}{2} g \mu_\text{B} B + \gamma_2 B^2 
\end{equation}
%=========================================
\noindent
with the exciton energy $E_\text{X}^0$ at $B=0$, the Bohr magneton $\mu_\text{B}$, the exciton Landé factor $g$ and the diamagnetic coefficient $\gamma_2$. The diamagnetic coefficient is proportional to the electron-hole coherence length $L_\text{eh}$ in the plane perpendicular to the magnetic field:\cite{Walck1998}

%=========================================
\begin{equation}\label{eq:gamma2}
\gamma_2 = \frac{e^2 L_\text{eh}}{8 \mu} 
\end{equation}
%=========================================

\begin{figure*}
\includegraphics[scale=1]{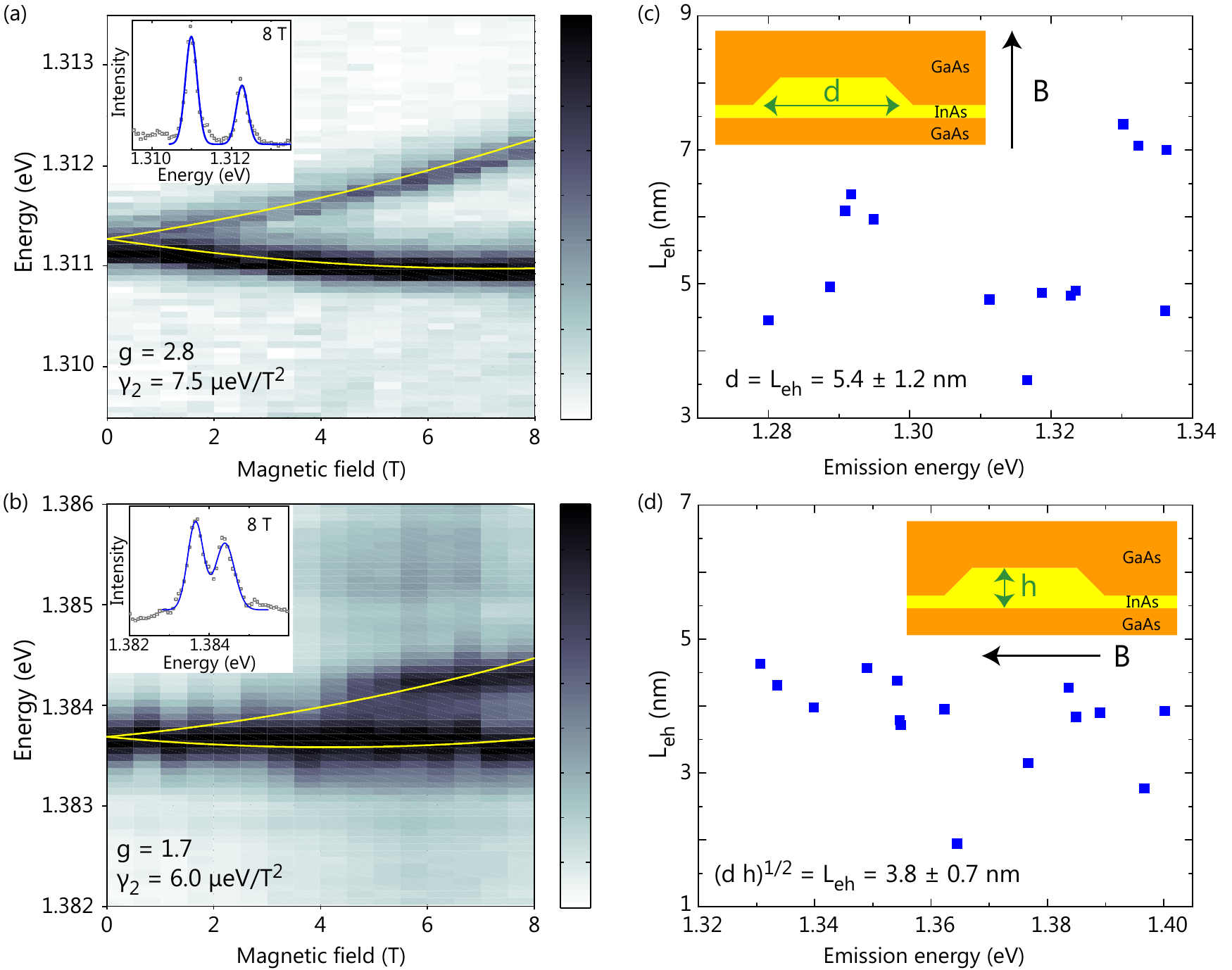}
\caption{(a) and (b) Evolution of the exciton energy $E_\text{X}$ in a magnetic field for a single InAs QD on (a) planar GaAs(110) and (b) a GaAs\{1$\bar{1}$0\} nanowire sidewall at 4.2~K. Both measurements were done in Faraday geometry. The intensities are color-coded according to the scale bars on the right. The solid lines are quadratic fits to the emission energies of the bright exciton states, yielding values for $g$ and $\gamma_2$ as indicated in the figures. The respective insets show PL spectra at 8~T (symbols) together with Gaussian fits (line) to determine the transition energies. (c) and (d) Electron-hole coherence length $L_\text{eh}$ for (c) 14 QDs grown on planar GaAs(110) and (d) 15 QDs on nanowire sidewalls. The mean values are indicated in the figure. The insets display a schematic representation of an InAs(110) QD in GaAs.}
\label{fig:Figure3}
\end{figure*} 

\noindent
with the reduced mass $\mu$ of the exciton in the plane perpendicular to the magnetic field. While large three-dimensional InAs islands on GaAs(110) are elongated along $[1\bar{1}0]$, possibly due to different adatom diffusivities along the $[1\bar{1}0]$ and $[001]$ directions,\cite{Ishii2003} the shape anisotropy for smaller QDs is negligible.\cite{Lewis2017a} If we suppose that, as depicted in the inset of Figs.~\ref{fig:Figure3}(c) and \ref{fig:Figure3}(d), the InAs(110) QDs in planar and nanowire samples are lens-shaped with a diameter $d$ and a height $h$, and that $L_\text{eh}$ for strongly confined excitons is given by the QD size, then $L_\text{eh} = d$ for InAs QDs grown on planar GaAs(110). For the nanowire sample measured in Faraday geometry, $\gamma_2$ is proportional to the coherence length of the exciton in the plane perpendicular to the nanowire axis. Therefore, $L_\text{eh}$ depends on both $d$ and $h$ and may be written approximately as $L_\text{eh} = \sqrt{d h}$. 

Figure~\ref{fig:Figure3}(a) shows the magnetic field dependence of the emission of a neutral exciton in an InAs QD on planar GaAs(110). For finite magnetic fields, the PL line is observed to split into two transitions, which we attribute to the two bright states of the exciton. QDs on GaAs(110) have $C_s$ symmetry, for which the magnetic field is expected to mix dark and bright exciton states,\cite{Bayer2002} hence giving rise to four distinct transitions. The fact that only two lines are observed even at 8~T [see inset in Fig.~\ref{fig:Figure3}(a)] suggests that the bright and dark exciton states have similar $g$ factors. In InAs QDs, the electron $g$ factor is known to depend only weakly on the QD size and shape \cite{Pryor2006} and to have a value between $-0.2$ and $-0.5$.\cite{Oberli2009,Schwan2011,VanHattem2013} A similar $g$ factor for bright and dark excitons thus implies a small hole $g$ factor for the InAs/GaAs(110) QDs under investigation. To obtain the energy of the exciton bright states as a function of $B$, the two lines observed in each PL spectrum are fit by Gaussians [see inset in Fig.~\ref{fig:Figure3}(a)]. The resulting transition energies are subsequently fit by Eq.~\ref{eq:shift} as shown by the solid lines in Fig.~\ref{fig:Figure3}(a), yielding $|g| = 2.8$ and $\gamma_2$ = 7.5~\textmu eV/T$^2$. Figure~\ref{fig:Figure3}(b) shows the electron-hole coherence lengths $L_\text{eh}$ deduced from the values of $\gamma_2$ measured for 14 different QDs emitting between 1.27 and 1.34~eV. As an average, we obtain $d = L_\text{eh} = (5.4 \pm 1.2)$~nm, a value similar to that reported for Stranski-Krastanov InAs/GaAs(001) QDs emitting in the same spectral range.\cite{VanHattem2013} 

The magnetic field dependence of the emission of a neutral exciton in an InAs QD on a \{110\} sidewall of a GaAs nanowire is shown in Fig.~\ref{fig:Figure3}(b). Similar to the planar sample in Fig.~\ref{fig:Figure3}(a), the PL line of the exciton in the sidewall InAs QDs splits into two transitions in the magnetic field. The energy of these lines follows a parabolic dependence on $B$ as well. Note that in contrast to the planar case, the PL lines at 8~T exhibit an asymmetric lineshape, which may be due to a contribution from the dark states. The $g$ factor and $\gamma_2$ values deduced from these experiments are smaller than those obtained for the planar sample [Fig.~\ref{fig:Figure3}(a)]. The former finding suggests that the exciton $g$ factor is anisotropic, most probably due to some anisotropy in the hole $g$ factor.\cite{Nakaoka2005,VanHattem2013} Measuring smaller $\gamma_2$ values for $B || [111]$ than for $B || [110]$ indicates that $h < d$. To confirm this result, we have measured $\gamma_2$ for 15 different InAs QDs on nanowire sidewalls. The corresponding values for $L_\text{eh}$ are shown in Fig.~\ref{fig:Figure3}(d), from which we arrive at an average $L_\text{eh} = (3.8 \pm 0.7)$~nm. Therefore, the strong confinement direction for these InAs(110) QDs is the [110] direction perpendicular to the surface. Assuming that $d$ is equal for QDs in the planar and nanowire samples, this result yields $h = 2.5$~nm, in good agreement with the result of the atom force microscopy study of uncapped QDs in Ref.~\onlinecite{Lewis2017a}.

\begin{figure*}
\includegraphics[scale=1]{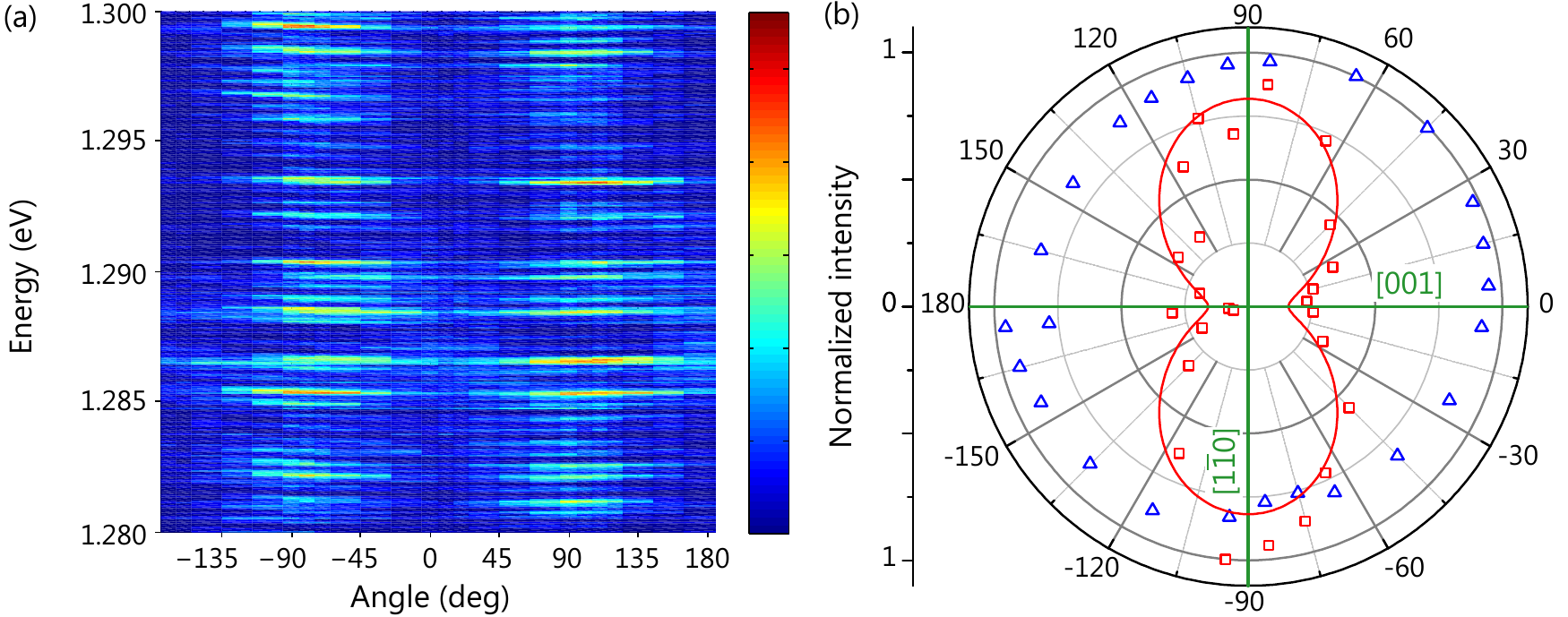}
\caption{(a) Polarization dependence of the transitions from neutral excitons confined in InAs QDs on planar GaAs(110) at 9~K. The intensities are color-coded according to the scale bar on the right. Angles of 0 and 90° correspond to light polarized along the $[001]$ and $[1\bar{1}0]$ directions, respectively. (b) Polar plot of the total intensity of the transitions between 1.28 and 1.30~eV (squares) as a function of the polarization angle. The solid line is a fit, indicating that the QD emission is polarized linearly along the $[1\bar{1}0]$ direction with a polarization ratio $\rho = -0.68$. The triangles represent the PL intensity from the GaAs substrate reflecting the polarization of the setup.}
\label{fig:Figure4}
\end{figure*} 

Together with the similar transition energies [Figs~\ref{fig:Figure2}(b,d)], the magneto-PL measurements in Fig~\ref{fig:Figure3} imply that the QDs in the planar and nanowire samples are of similar shape and size. Hence, we suggest that the different character of the biexciton state in these samples is not only related to the size of the QD, but also to the strain state. In fact, strain results in piezoelectric fields in the (110) plane of the QD \cite{Singh2013} and, depending on the magnitude of these built-in fields, biexcitons in (In,Ga)As(110) QDs can be binding or antibinding.\cite{Simeonov2008} Moreover, it has been predicted that the exact strain state in an InAs QD on a \{1$\bar{1}$0\} nanowire sidewall differs from that of a QD on planar GaAs(110). In the former case, the substrate is rigid and both the WL and the QD base are in perfect registry with the unstrained GaAs lattice. In the latter, however, the nanowire geometry results in enhanced elastic strain relaxation in all three dimensions. It is thus expected that the strain state of QDs in nanowires will differ from that of QDs on planar GaAs(110).\citep{Salehzadeh2013} In particular, since the core and shell materials assume the same lattice constant along the nanowire axis, they both experience a uniaxial strain. Since the NW strain state depends on the entire structure, tuning the strain by adjusting the nanowire core-shell structure could provide a way to tune the biexciton binding energy. This feature could, for instance, allow one to fabricate QDs with zero biexciton binding energy, which would be suitable for the generation of pairs of entangled photons.\cite{Avron2008}

As mentioned in the introduction, the growth of InAs QDs on GaAs(110) rather than on GaAs(001) has profound consequences resulting from the inequivalent $[1\bar{1}0]$ and $[001]$ in-plane directions of InAs(110). In contrast to the $C_{2v}$ symmetry of InAs QDs on GaAs(001), these inequivalent directions result in a $C_s$ symmetry with the (1$\bar{1}$0) plane acting as a reflection plane.\cite{Singh2013} Theoretically, the bright exciton states in InAs(110) QDs are expected to exhibit a comparatively large anisotropic exchange splitting, and the corresponding transitions should be polarized along the $[1\bar{1}0]$ and $[001]$ directions.\cite{Singh2013} Figure~\ref{fig:Figure4}(a) shows the polarization dependence of the transitions from several QDs on planar GaAs(110)  between 1.28 and 1.30~eV. This measurement was carried out at zero magnetic field and with an excitation power low enough to ensure that most transitions arise from the recombination of neutral excitons. An angle of 0° corresponds to light polarized along $[001]$. Evidently, while intense QD transitions are observed for light polarized along $[1\bar{1}0]$, the signal is too weak for light polarized along $[001]$ to allow a reliable measurement of the anisotropic exchange splitting between the two bright exciton states. We note that such a measurement would be even more complex for QDs in nanowires, since the nanowire geometry leads to an antenna effect that results in different extraction efficiencies for light polarized along the  $[1\bar{1}0]$ and $[001]$ directions.\cite{Fontana2014}

The strong intensity difference between the bright exciton states for $[1\bar{1}0]$ and $[001]$ polarization visible in Fig.~\ref{fig:Figure4}(a) could arise from an anisotropic exchange splitting for InAs(110) QDs that is so large that only the state at lower energy is occupied at low temperatures.\cite{Honig2014} This explanation can be safely excluded here. As shown in Fig.~\ref{fig:Figure3}(a), the energies of the exciton bright states for the planar sample follow a clear parabolic dependence for fields larger than about 2~T. In other words, the Zeeman splitting of the exciton bright states at 2~T is much larger than their anisotropic exchange splitting. Using $|g| = 2.8$ [Fig.~\ref{fig:Figure3}(a)], we estimate that the anisotropic exchange splitting for InAs(110) QDs is smaller than 100~\textmu eV.

Using atomistic calculations, \textcite{Singh2013} showed that the strain in InAs(110) QDs may lead to a mixing between heavy-hole and light-hole states. In contrast to InAs(001) QDs with a $C_{2v}$ symmetry, the mixing is strong for the [110] orientation since heavy and light holes belong to the same irreducible representation for the $C_s$ point group. As a result of this mixing, the PL signal from excitons in InAs(110) QDs may be linearly polarized. To obtain the polarization degree for the bright excitons in our InAs QDs on planar GaAs(110), we plot in Fig.~\ref{fig:Figure4}(b) the polarization dependence of the intensity summed over all QDs emitting between 1.28 and 1.30~eV in Fig.~\ref{fig:Figure4}(a). For comparison, we also show the polarization dependence of the near band edge emission from the GaAs substrate, which should be entirely unpolarized\cite{Kajikawa1991}. Evidently, the polarization dependence of our setup can be neglected for the present analysis. The QD PL signal is clearly linearly polarized along [1$\bar{1}$0], in agreement with the data in Fig.~\ref{fig:Figure4}(a). With the degree of linear polarization $\rho = (I_{001} - I_{1\bar{1}0})/(I_{001} + I_{1\bar{1}0})$, where $I_{001}$ and $I_{1\bar{1}0}$ are the PL intensities for light polarized along [001] and [1$\bar{1}$0], respectively, we obtain $\rho = -0.68$ as the average polarization for QDs emitting between 1.28 and 1.30~eV. This polarization degree is not only opposite in sign compared to that computed for InAs QDs on GaAs(110) by \citet{Singh2013} ($\rho = 0.35$), but also significantly larger. Apparently, the much smaller base diameter of our InAs QDs [cf.\ Fig.~\ref{fig:Figure3}(b)] in comparison to the 25~nm diameter considered in Ref.~\onlinecite{Singh2013} enhances the heavy-hole/light-hole mixing and thus amplifies the associated polarization of the bright exciton state. In any case, the findings in Fig.~\ref{fig:Figure4} demonstrate the possibility of fabricating linearly polarized single photon emitters with a deterministic polarization axis determined by the crystallographic axes.

\section{Summary and Conclusions}

We have studied the fine structure of excitons confined in InAs(110) QDs grown by molecular beam epitaxy. Employing a morphological instability induced by the surfactant Bi, these QDs form on planar GaAs(110) as well as on the $\{110\}$ sidewall facets of GaAs nanowires. Light emission associated with the radiative decay of excitons in the QDs has been observed in PL spectra between 1.1 and 1.42~eV. From magneto-PL experiments, we have shown that the strong confinement axis is the [110] direction normal to the surface, and we have estimated that the smallest QDs have a height of about 2.5~nm. Despite their reduced symmetry compared to InAs QDs grown on (001) or (111) surfaces, these QDs constitute a versatile system for quantum optics applications in the near infrared spectral range. First, the binding energy of biexcitons in (110) InAs QDs not only depends on the dimensions of the QD, but also on the strength of the piezoelectric fields in the (110) plane. We have observed both binding and antibinding biexcitons, suggesting that one could produce QDs with zero biexciton binding energy by tuning the QD shape and strain state. As a practical means for tuning the strain, we propose to vary the thickness of the GaAs core and outer-shell of GaAs/InAs core-multishells nanowires. Alternatively, variation in the biexciton energy could be achieved by mechanically driving the nanowire. Despite a nonzero anisotropic exchange splitting, such QDs could be used as entangled photon pairs emitter via the time reordering scheme.\cite{Avron2008} Furthermore, the strain in the (110) plane mixes the heavy and light holes states. As a result of this mixing, the photoluminescence signal of InAs QDs is polarized along the $[1\bar{1}0]$ or the [001] direction, depending on the exact shape and In content of the QD. The high degree of linear polarization of InAs QDs along a well-defined axis opens the possibility to generate linearly polarized single photons, which is of interest for applications in quantum key distribution.\cite{Muller1993}

\acknowledgments
We are grateful to Gerd Paris and Manfred Ramsteiner for technical assistance with the confocal optical setup, and to Jesús Herranz for a careful reading of the manuscript. P.\,C. acknowledges funding from the Fonds National Suisse de la Recherche Scientifique through project No.~161032. R.\,B.\,L. acknowledges support from the Alexander von Humboldt foundation.

\bibliography{bibliography}

%merlin.mbs apsrev4-1.bst 2010-07-25 4.21a (PWD, AO, DPC) hacked
%Control: key (0)
%Control: author (8) initials jnrlst
%Control: editor formatted (1) identically to author
%Control: production of article title (-1) disabled
%Control: page (0) single
%Control: year (1) truncated
%Control: production of eprint (0) enabled
\begin{thebibliography}{39}%
\makeatletter
\providecommand \@ifxundefined [1]{%
 \@ifx{#1\undefined}
}%
\providecommand \@ifnum [1]{%
 \ifnum #1\expandafter \@firstoftwo
 \else \expandafter \@secondoftwo
 \fi
}%
\providecommand \@ifx [1]{%
 \ifx #1\expandafter \@firstoftwo
 \else \expandafter \@secondoftwo
 \fi
}%
\providecommand \natexlab [1]{#1}%
\providecommand \enquote  [1]{``#1''}%
\providecommand \bibnamefont  [1]{#1}%
\providecommand \bibfnamefont [1]{#1}%
\providecommand \citenamefont [1]{#1}%
\providecommand \href@noop [0]{\@secondoftwo}%
\providecommand \href [0]{\begingroup \@sanitize@url \@href}%
\providecommand \@href[1]{\@@startlink{#1}\@@href}%
\providecommand \@@href[1]{\endgroup#1\@@endlink}%
\providecommand \@sanitize@url [0]{\catcode `\\12\catcode `\$12\catcode
  `\&12\catcode `\#12\catcode `\^12\catcode `\_12\catcode `\%12\relax}%
\providecommand \@@startlink[1]{}%
\providecommand \@@endlink[0]{}%
\providecommand \url  [0]{\begingroup\@sanitize@url \@url }%
\providecommand \@url [1]{\endgroup\@href {#1}{\urlprefix }}%
\providecommand \urlprefix  [0]{URL }%
\providecommand \Eprint [0]{\href }%
\providecommand \doibase [0]{http://dx.doi.org/}%
\providecommand \selectlanguage [0]{\@gobble}%
\providecommand \bibinfo  [0]{\@secondoftwo}%
\providecommand \bibfield  [0]{\@secondoftwo}%
\providecommand \translation [1]{[#1]}%
\providecommand \BibitemOpen [0]{}%
\providecommand \bibitemStop [0]{}%
\providecommand \bibitemNoStop [0]{.\EOS\space}%
\providecommand \EOS [0]{\spacefactor3000\relax}%
\providecommand \BibitemShut  [1]{\csname bibitem#1\endcsname}%
\let\auto@bib@innerbib\@empty
%</preamble>
\bibitem [{\citenamefont {Michler}\ \emph {et~al.}(2000)\citenamefont
  {Michler}, \citenamefont {Kiraz}, \citenamefont {Becher}, \citenamefont
  {Schoenfeld}, \citenamefont {Petroff}, \citenamefont {Zhang}, \citenamefont
  {Hu},\ and\ \citenamefont {Imamoglu}}]{Michler2000}%
  \BibitemOpen
  \bibfield  {author} {\bibinfo {author} {\bibfnamefont {P.}~\bibnamefont
  {Michler}}, \bibinfo {author} {\bibfnamefont {A.}~\bibnamefont {Kiraz}},
  \bibinfo {author} {\bibfnamefont {C.}~\bibnamefont {Becher}}, \bibinfo
  {author} {\bibfnamefont {W.~V.}\ \bibnamefont {Schoenfeld}}, \bibinfo
  {author} {\bibfnamefont {P.~M.}\ \bibnamefont {Petroff}}, \bibinfo {author}
  {\bibfnamefont {L.}~\bibnamefont {Zhang}}, \bibinfo {author} {\bibfnamefont
  {E.}~\bibnamefont {Hu}}, \ and\ \bibinfo {author} {\bibfnamefont
  {A.}~\bibnamefont {Imamoglu}},\ }\href {\doibase
  10.1126/science.290.5500.2282} {\bibfield  {journal} {\bibinfo  {journal}
  {Science}\ }\textbf {\bibinfo {volume} {290}},\ \bibinfo {pages} {2282}
  (\bibinfo {year} {2000})}\BibitemShut {NoStop}%
\bibitem [{\citenamefont {Akopian}\ \emph {et~al.}(2006)\citenamefont
  {Akopian}, \citenamefont {Lindner}, \citenamefont {Poem}, \citenamefont
  {Berlatzky}, \citenamefont {Avron}, \citenamefont {Gershoni}, \citenamefont
  {Gerardot},\ and\ \citenamefont {Petroff}}]{Akopian2006}%
  \BibitemOpen
  \bibfield  {author} {\bibinfo {author} {\bibfnamefont {N.}~\bibnamefont
  {Akopian}}, \bibinfo {author} {\bibfnamefont {N.~H.}\ \bibnamefont
  {Lindner}}, \bibinfo {author} {\bibfnamefont {E.}~\bibnamefont {Poem}},
  \bibinfo {author} {\bibfnamefont {Y.}~\bibnamefont {Berlatzky}}, \bibinfo
  {author} {\bibfnamefont {J.}~\bibnamefont {Avron}}, \bibinfo {author}
  {\bibfnamefont {D.}~\bibnamefont {Gershoni}}, \bibinfo {author}
  {\bibfnamefont {B.~D.}\ \bibnamefont {Gerardot}}, \ and\ \bibinfo {author}
  {\bibfnamefont {P.~M.}\ \bibnamefont {Petroff}},\ }\href {\doibase
  10.1103/PhysRevLett.96.130501} {\bibfield  {journal} {\bibinfo  {journal}
  {Phys. Rev. Lett.}\ }\textbf {\bibinfo {volume} {96}},\ \bibinfo {pages}
  {130501} (\bibinfo {year} {2006})}\BibitemShut {NoStop}%
\bibitem [{\citenamefont {Stevenson}\ \emph {et~al.}(2006)\citenamefont
  {Stevenson}, \citenamefont {Young}, \citenamefont {Atkinson}, \citenamefont
  {Cooper}, \citenamefont {Ritchie},\ and\ \citenamefont
  {Shields}}]{Stevenson2006}%
  \BibitemOpen
  \bibfield  {author} {\bibinfo {author} {\bibfnamefont {R.~M.}\ \bibnamefont
  {Stevenson}}, \bibinfo {author} {\bibfnamefont {R.~J.}\ \bibnamefont
  {Young}}, \bibinfo {author} {\bibfnamefont {P.}~\bibnamefont {Atkinson}},
  \bibinfo {author} {\bibfnamefont {K.}~\bibnamefont {Cooper}}, \bibinfo
  {author} {\bibfnamefont {D.~A.}\ \bibnamefont {Ritchie}}, \ and\ \bibinfo
  {author} {\bibfnamefont {A.~J.}\ \bibnamefont {Shields}},\ }\href {\doibase
  10.1038/nature04446} {\bibfield  {journal} {\bibinfo  {journal} {Nature}\
  }\textbf {\bibinfo {volume} {439}},\ \bibinfo {pages} {179} (\bibinfo {year}
  {2006})}\BibitemShut {NoStop}%
\bibitem [{\citenamefont {Li}\ \emph {et~al.}(2003)\citenamefont {Li},
  \citenamefont {Wu}, \citenamefont {Steel}, \citenamefont {Gammon},
  \citenamefont {Stievater}, \citenamefont {Katzer}, \citenamefont {Park},
  \citenamefont {Piermarocchi},\ and\ \citenamefont {Sham}}]{Li2003}%
  \BibitemOpen
  \bibfield  {author} {\bibinfo {author} {\bibfnamefont {X.}~\bibnamefont
  {Li}}, \bibinfo {author} {\bibfnamefont {Y.}~\bibnamefont {Wu}}, \bibinfo
  {author} {\bibfnamefont {D.}~\bibnamefont {Steel}}, \bibinfo {author}
  {\bibfnamefont {D.}~\bibnamefont {Gammon}}, \bibinfo {author} {\bibfnamefont
  {T.~H.}\ \bibnamefont {Stievater}}, \bibinfo {author} {\bibfnamefont {D.~S.}\
  \bibnamefont {Katzer}}, \bibinfo {author} {\bibfnamefont {D.}~\bibnamefont
  {Park}}, \bibinfo {author} {\bibfnamefont {C.}~\bibnamefont {Piermarocchi}},
  \ and\ \bibinfo {author} {\bibfnamefont {L.~J.}\ \bibnamefont {Sham}},\
  }\href {\doibase 10.1126/science.1083800} {\bibfield  {journal} {\bibinfo
  {journal} {Science}\ }\textbf {\bibinfo {volume} {301}},\ \bibinfo {pages}
  {809} (\bibinfo {year} {2003})}\BibitemShut {NoStop}%
\bibitem [{\citenamefont {Benson}\ \emph {et~al.}(2000)\citenamefont {Benson},
  \citenamefont {Santori}, \citenamefont {Pelton},\ and\ \citenamefont
  {Yamamoto}}]{Benson2000}%
  \BibitemOpen
  \bibfield  {author} {\bibinfo {author} {\bibfnamefont {O.}~\bibnamefont
  {Benson}}, \bibinfo {author} {\bibfnamefont {C.}~\bibnamefont {Santori}},
  \bibinfo {author} {\bibfnamefont {M.}~\bibnamefont {Pelton}}, \ and\ \bibinfo
  {author} {\bibfnamefont {Y.}~\bibnamefont {Yamamoto}},\ }\href {\doibase
  10.1103/PhysRevLett.84.2513} {\bibfield  {journal} {\bibinfo  {journal}
  {Phys. Rev. Lett.}\ }\textbf {\bibinfo {volume} {84}},\ \bibinfo {pages}
  {2513} (\bibinfo {year} {2000})}\BibitemShut {NoStop}%
\bibitem [{\citenamefont {Schliwa}\ \emph {et~al.}(2009)\citenamefont
  {Schliwa}, \citenamefont {Winkelnkemper}, \citenamefont {Lochmann},
  \citenamefont {Stock},\ and\ \citenamefont {Bimberg}}]{Schliwa2009}%
  \BibitemOpen
  \bibfield  {author} {\bibinfo {author} {\bibfnamefont {A.}~\bibnamefont
  {Schliwa}}, \bibinfo {author} {\bibfnamefont {M.}~\bibnamefont
  {Winkelnkemper}}, \bibinfo {author} {\bibfnamefont {A.}~\bibnamefont
  {Lochmann}}, \bibinfo {author} {\bibfnamefont {E.}~\bibnamefont {Stock}}, \
  and\ \bibinfo {author} {\bibfnamefont {D.}~\bibnamefont {Bimberg}},\ }\href
  {\doibase 10.1103/PhysRevB.80.161307} {\bibfield  {journal} {\bibinfo
  {journal} {Phys. Rev. B}\ }\textbf {\bibinfo {volume} {80}},\ \bibinfo
  {pages} {161307} (\bibinfo {year} {2009})}\BibitemShut {NoStop}%
\bibitem [{\citenamefont {Bayer}\ \emph {et~al.}(2002)\citenamefont {Bayer},
  \citenamefont {Ortner}, \citenamefont {Stern}, \citenamefont {Kuther},
  \citenamefont {Gorbunov}, \citenamefont {Forchel}, \citenamefont {Hawrylak},
  \citenamefont {Fafard}, \citenamefont {Hinzer}, \citenamefont {Reinecke},
  \citenamefont {Walck}, \citenamefont {Reithmaier}, \citenamefont {Klopf},\
  and\ \citenamefont {Sch\"afer}}]{Bayer2002}%
  \BibitemOpen
  \bibfield  {author} {\bibinfo {author} {\bibfnamefont {M.}~\bibnamefont
  {Bayer}}, \bibinfo {author} {\bibfnamefont {G.}~\bibnamefont {Ortner}},
  \bibinfo {author} {\bibfnamefont {O.}~\bibnamefont {Stern}}, \bibinfo
  {author} {\bibfnamefont {A.}~\bibnamefont {Kuther}}, \bibinfo {author}
  {\bibfnamefont {A.~A.}\ \bibnamefont {Gorbunov}}, \bibinfo {author}
  {\bibfnamefont {A.}~\bibnamefont {Forchel}}, \bibinfo {author} {\bibfnamefont
  {P.}~\bibnamefont {Hawrylak}}, \bibinfo {author} {\bibfnamefont
  {S.}~\bibnamefont {Fafard}}, \bibinfo {author} {\bibfnamefont
  {K.}~\bibnamefont {Hinzer}}, \bibinfo {author} {\bibfnamefont {T.~L.}\
  \bibnamefont {Reinecke}}, \bibinfo {author} {\bibfnamefont {S.~N.}\
  \bibnamefont {Walck}}, \bibinfo {author} {\bibfnamefont {J.~P.}\ \bibnamefont
  {Reithmaier}}, \bibinfo {author} {\bibfnamefont {F.}~\bibnamefont {Klopf}}, \
  and\ \bibinfo {author} {\bibfnamefont {F.}~\bibnamefont {Sch\"afer}},\ }\href
  {\doibase 10.1103/PhysRevB.65.195315} {\bibfield  {journal} {\bibinfo
  {journal} {Phys. Rev. B}\ }\textbf {\bibinfo {volume} {65}},\ \bibinfo
  {pages} {195315} (\bibinfo {year} {2002})}\BibitemShut {NoStop}%
\bibitem [{\citenamefont {Juska}\ \emph {et~al.}(2013)\citenamefont {Juska},
  \citenamefont {Dimastrodonato}, \citenamefont {Mereni}, \citenamefont
  {Gocalinska},\ and\ \citenamefont {Pelucchi}}]{Juska2013}%
  \BibitemOpen
  \bibfield  {author} {\bibinfo {author} {\bibfnamefont {G.}~\bibnamefont
  {Juska}}, \bibinfo {author} {\bibfnamefont {V.}~\bibnamefont
  {Dimastrodonato}}, \bibinfo {author} {\bibfnamefont {L.~O.}\ \bibnamefont
  {Mereni}}, \bibinfo {author} {\bibfnamefont {A.}~\bibnamefont {Gocalinska}},
  \ and\ \bibinfo {author} {\bibfnamefont {E.}~\bibnamefont {Pelucchi}},\
  }\href {\doibase 10.1038/nphoton.2013.128} {\bibfield  {journal} {\bibinfo
  {journal} {Nat. Photon.}\ }\textbf {\bibinfo {volume} {7}},\ \bibinfo {pages}
  {527} (\bibinfo {year} {2013})}\BibitemShut {NoStop}%
\bibitem [{\citenamefont {Juska}\ \emph {et~al.}(2014)\citenamefont {Juska},
  \citenamefont {Dimastrodonato}, \citenamefont {Mereni}, \citenamefont
  {Chung}, \citenamefont {Gocalinska}, \citenamefont {Pelucchi}, \citenamefont
  {Van~Hattem}, \citenamefont {Ediger},\ and\ \citenamefont
  {Corfdir}}]{Juska2014}%
  \BibitemOpen
  \bibfield  {author} {\bibinfo {author} {\bibfnamefont {G.}~\bibnamefont
  {Juska}}, \bibinfo {author} {\bibfnamefont {V.}~\bibnamefont
  {Dimastrodonato}}, \bibinfo {author} {\bibfnamefont {L.~O.}\ \bibnamefont
  {Mereni}}, \bibinfo {author} {\bibfnamefont {T.~H.}\ \bibnamefont {Chung}},
  \bibinfo {author} {\bibfnamefont {A.}~\bibnamefont {Gocalinska}}, \bibinfo
  {author} {\bibfnamefont {E.}~\bibnamefont {Pelucchi}}, \bibinfo {author}
  {\bibfnamefont {B.}~\bibnamefont {Van~Hattem}}, \bibinfo {author}
  {\bibfnamefont {M.}~\bibnamefont {Ediger}}, \ and\ \bibinfo {author}
  {\bibfnamefont {P.}~\bibnamefont {Corfdir}},\ }\href {\doibase
  10.1103/PhysRevB.89.205430} {\bibfield  {journal} {\bibinfo  {journal} {Phys.
  Rev. B}\ }\textbf {\bibinfo {volume} {89}},\ \bibinfo {pages} {205430}
  (\bibinfo {year} {2014})}\BibitemShut {NoStop}%
\bibitem [{\citenamefont {Mano}\ \emph {et~al.}(2008)\citenamefont {Mano},
  \citenamefont {Kuroda}, \citenamefont {Mitsuishi}, \citenamefont {Nakayama},
  \citenamefont {Noda},\ and\ \citenamefont {Sakoda}}]{Mano2008}%
  \BibitemOpen
  \bibfield  {author} {\bibinfo {author} {\bibfnamefont {T.}~\bibnamefont
  {Mano}}, \bibinfo {author} {\bibfnamefont {T.}~\bibnamefont {Kuroda}},
  \bibinfo {author} {\bibfnamefont {K.}~\bibnamefont {Mitsuishi}}, \bibinfo
  {author} {\bibfnamefont {Y.}~\bibnamefont {Nakayama}}, \bibinfo {author}
  {\bibfnamefont {T.}~\bibnamefont {Noda}}, \ and\ \bibinfo {author}
  {\bibfnamefont {K.}~\bibnamefont {Sakoda}},\ }\href {\doibase
  10.1063/1.3026174} {\bibfield  {journal} {\bibinfo  {journal} {Appl. Phys.
  Lett.}\ }\textbf {\bibinfo {volume} {93}},\ \bibinfo {pages} {203110}
  (\bibinfo {year} {2008})}\BibitemShut {NoStop}%
\bibitem [{\citenamefont {Wen}\ \emph {et~al.}(2016)\citenamefont {Wen},
  \citenamefont {Gao}, \citenamefont {Zhang},\ and\ \citenamefont
  {Li}}]{Lei2016}%
  \BibitemOpen
  \bibfield  {author} {\bibinfo {author} {\bibfnamefont {L.}~\bibnamefont
  {Wen}}, \bibinfo {author} {\bibfnamefont {F.}~\bibnamefont {Gao}}, \bibinfo
  {author} {\bibfnamefont {S.}~\bibnamefont {Zhang}}, \ and\ \bibinfo {author}
  {\bibfnamefont {G.}~\bibnamefont {Li}},\ }\href {\doibase
  10.1002/smll.201503387} {\bibfield  {journal} {\bibinfo  {journal} {Small}\
  }\textbf {\bibinfo {volume} {12}},\ \bibinfo {pages} {4277} (\bibinfo {year}
  {2016})}\BibitemShut {NoStop}%
\bibitem [{\citenamefont {{Lewis}}\ \emph
  {et~al.}(2017{\natexlab{a}})\citenamefont {{Lewis}}, \citenamefont
  {{Corfdir}}, \citenamefont {{Li}}, \citenamefont {{Herranz}}, \citenamefont
  {{Pf{\"u}ller}}, \citenamefont {{Brandt}},\ and\ \citenamefont
  {{Geelhaar}}}]{Lewis2017a}%
  \BibitemOpen
  \bibfield  {author} {\bibinfo {author} {\bibfnamefont {R.~B.}\ \bibnamefont
  {{Lewis}}}, \bibinfo {author} {\bibfnamefont {P.}~\bibnamefont {{Corfdir}}},
  \bibinfo {author} {\bibfnamefont {H.}~\bibnamefont {{Li}}}, \bibinfo {author}
  {\bibfnamefont {J.}~\bibnamefont {{Herranz}}}, \bibinfo {author}
  {\bibfnamefont {C.}~\bibnamefont {{Pf{\"u}ller}}}, \bibinfo {author}
  {\bibfnamefont {O.}~\bibnamefont {{Brandt}}}, \ and\ \bibinfo {author}
  {\bibfnamefont {L.}~\bibnamefont {{Geelhaar}}},\ }\href@noop {} {\bibfield
  {journal} {\bibinfo  {journal} {ArXiv e-prints}\ } (\bibinfo {year}
  {2017}{\natexlab{a}})},\ \Eprint {http://arxiv.org/abs/1703.05025}
  {arXiv:1703.05025 [cond-mat.mtrl-sci]} \BibitemShut {NoStop}%
\bibitem [{\citenamefont {{Lewis}}\ \emph
  {et~al.}(2017{\natexlab{b}})\citenamefont {{Lewis}}, \citenamefont
  {{Corfdir}}, \citenamefont {{Herranz}}, \citenamefont {{K{\"u}pers}},
  \citenamefont {{Jahn}}, \citenamefont {{Brandt}},\ and\ \citenamefont
  {{Geelhaar}}}]{Lewis2017b}%
  \BibitemOpen
  \bibfield  {author} {\bibinfo {author} {\bibfnamefont {R.~B.}\ \bibnamefont
  {{Lewis}}}, \bibinfo {author} {\bibfnamefont {P.}~\bibnamefont {{Corfdir}}},
  \bibinfo {author} {\bibfnamefont {J.}~\bibnamefont {{Herranz}}}, \bibinfo
  {author} {\bibfnamefont {H.}~\bibnamefont {{K{\"u}pers}}}, \bibinfo {author}
  {\bibfnamefont {U.}~\bibnamefont {{Jahn}}}, \bibinfo {author} {\bibfnamefont
  {O.}~\bibnamefont {{Brandt}}}, \ and\ \bibinfo {author} {\bibfnamefont
  {L.}~\bibnamefont {{Geelhaar}}},\ }\href@noop {} {\bibfield  {journal}
  {\bibinfo  {journal} {ArXiv e-prints}\ } (\bibinfo {year}
  {2017}{\natexlab{b}})},\ \Eprint {http://arxiv.org/abs/1704.08014}
  {arXiv:1704.08014 [cond-mat.mtrl-sci]} \BibitemShut {NoStop}%
\bibitem [{\citenamefont {Singh}\ and\ \citenamefont
  {Bester}(2013)}]{Singh2013}%
  \BibitemOpen
  \bibfield  {author} {\bibinfo {author} {\bibfnamefont {R.}~\bibnamefont
  {Singh}}\ and\ \bibinfo {author} {\bibfnamefont {G.}~\bibnamefont {Bester}},\
  }\href {\doibase 10.1103/PhysRevB.88.075430} {\bibfield  {journal} {\bibinfo
  {journal} {Phys. Rev. B}\ }\textbf {\bibinfo {volume} {88}},\ \bibinfo
  {pages} {075430} (\bibinfo {year} {2013})}\BibitemShut {NoStop}%
\bibitem [{\citenamefont {Belk}\ \emph {et~al.}(1997)\citenamefont {Belk},
  \citenamefont {Sudijono}, \citenamefont {Zhang}, \citenamefont {Neave},
  \citenamefont {Jones},\ and\ \citenamefont {Joyce}}]{Belk1997}%
  \BibitemOpen
  \bibfield  {author} {\bibinfo {author} {\bibfnamefont {J.~G.}\ \bibnamefont
  {Belk}}, \bibinfo {author} {\bibfnamefont {J.~L.}\ \bibnamefont {Sudijono}},
  \bibinfo {author} {\bibfnamefont {X.~M.}\ \bibnamefont {Zhang}}, \bibinfo
  {author} {\bibfnamefont {J.~H.}\ \bibnamefont {Neave}}, \bibinfo {author}
  {\bibfnamefont {T.~S.}\ \bibnamefont {Jones}}, \ and\ \bibinfo {author}
  {\bibfnamefont {B.~A.}\ \bibnamefont {Joyce}},\ }\href {\doibase
  10.1103/PhysRevLett.78.475} {\bibfield  {journal} {\bibinfo  {journal} {Phys.
  Rev. Lett.}\ }\textbf {\bibinfo {volume} {78}},\ \bibinfo {pages} {475}
  (\bibinfo {year} {1997})}\BibitemShut {NoStop}%
\bibitem [{\citenamefont {Corfdir}\ \emph
  {et~al.}(2016{\natexlab{a}})\citenamefont {Corfdir}, \citenamefont {Lewis},
  \citenamefont {Marquardt}, \citenamefont {K{\"{u}}pers}, \citenamefont
  {Grandal}, \citenamefont {Dimakis}, \citenamefont {Trampert}, \citenamefont
  {Geelhaar}, \citenamefont {Brandt},\ and\ \citenamefont
  {Phillips}}]{Corfdir2016b}%
  \BibitemOpen
  \bibfield  {author} {\bibinfo {author} {\bibfnamefont {P.}~\bibnamefont
  {Corfdir}}, \bibinfo {author} {\bibfnamefont {R.~B.}\ \bibnamefont {Lewis}},
  \bibinfo {author} {\bibfnamefont {O.}~\bibnamefont {Marquardt}}, \bibinfo
  {author} {\bibfnamefont {H.}~\bibnamefont {K{\"{u}}pers}}, \bibinfo {author}
  {\bibfnamefont {J.}~\bibnamefont {Grandal}}, \bibinfo {author} {\bibfnamefont
  {E.}~\bibnamefont {Dimakis}}, \bibinfo {author} {\bibfnamefont
  {A.}~\bibnamefont {Trampert}}, \bibinfo {author} {\bibfnamefont
  {L.}~\bibnamefont {Geelhaar}}, \bibinfo {author} {\bibfnamefont
  {O.}~\bibnamefont {Brandt}}, \ and\ \bibinfo {author} {\bibfnamefont {R.~T.}\
  \bibnamefont {Phillips}},\ }\href {\doibase 10.1063/1.4961245} {\bibfield
  {journal} {\bibinfo  {journal} {Appl. Phys. Lett.}\ }\textbf {\bibinfo
  {volume} {109}},\ \bibinfo {pages} {082107} (\bibinfo {year}
  {2016}{\natexlab{a}})}\BibitemShut {NoStop}%
\bibitem [{\citenamefont {Fontana}\ \emph {et~al.}(2014)\citenamefont
  {Fontana}, \citenamefont {Corfdir}, \citenamefont {Van~Hattem}, \citenamefont
  {Russo-Averchi}, \citenamefont {Heiss}, \citenamefont {Sonderegger},
  \citenamefont {Magen}, \citenamefont {Arbiol}, \citenamefont {Phillips},\
  and\ \citenamefont {Fontcuberta~i Morral}}]{Fontana2014}%
  \BibitemOpen
  \bibfield  {author} {\bibinfo {author} {\bibfnamefont {Y.}~\bibnamefont
  {Fontana}}, \bibinfo {author} {\bibfnamefont {P.}~\bibnamefont {Corfdir}},
  \bibinfo {author} {\bibfnamefont {B.}~\bibnamefont {Van~Hattem}}, \bibinfo
  {author} {\bibfnamefont {E.}~\bibnamefont {Russo-Averchi}}, \bibinfo {author}
  {\bibfnamefont {M.}~\bibnamefont {Heiss}}, \bibinfo {author} {\bibfnamefont
  {S.}~\bibnamefont {Sonderegger}}, \bibinfo {author} {\bibfnamefont
  {C.}~\bibnamefont {Magen}}, \bibinfo {author} {\bibfnamefont
  {J.}~\bibnamefont {Arbiol}}, \bibinfo {author} {\bibfnamefont {R.~T.}\
  \bibnamefont {Phillips}}, \ and\ \bibinfo {author} {\bibfnamefont
  {A.}~\bibnamefont {Fontcuberta~i Morral}},\ }\href {\doibase
  10.1103/PhysRevB.90.075307} {\bibfield  {journal} {\bibinfo  {journal} {Phys.
  Rev. B}\ }\textbf {\bibinfo {volume} {90}},\ \bibinfo {pages} {075307}
  (\bibinfo {year} {2014})}\BibitemShut {NoStop}%
\bibitem [{\citenamefont {Holmes}\ \emph {et~al.}(2015)\citenamefont {Holmes},
  \citenamefont {Kako}, \citenamefont {Choi}, \citenamefont {Arita},\ and\
  \citenamefont {Arakawa}}]{Holmes2015}%
  \BibitemOpen
  \bibfield  {author} {\bibinfo {author} {\bibfnamefont {M.}~\bibnamefont
  {Holmes}}, \bibinfo {author} {\bibfnamefont {S.}~\bibnamefont {Kako}},
  \bibinfo {author} {\bibfnamefont {K.}~\bibnamefont {Choi}}, \bibinfo {author}
  {\bibfnamefont {M.}~\bibnamefont {Arita}}, \ and\ \bibinfo {author}
  {\bibfnamefont {Y.}~\bibnamefont {Arakawa}},\ }\href {\doibase
  10.1103/PhysRevB.92.115447} {\bibfield  {journal} {\bibinfo  {journal} {Phys.
  Rev. B}\ }\textbf {\bibinfo {volume} {92}},\ \bibinfo {pages} {115447}
  (\bibinfo {year} {2015})}\BibitemShut {NoStop}%
\bibitem [{\citenamefont {Corfdir}\ \emph
  {et~al.}(2016{\natexlab{b}})\citenamefont {Corfdir}, \citenamefont
  {K{\"{u}}pers}, \citenamefont {Lewis}, \citenamefont {Flissikowski},
  \citenamefont {Grahn}, \citenamefont {Geelhaar},\ and\ \citenamefont
  {Brandt}}]{Corfdir2016}%
  \BibitemOpen
  \bibfield  {author} {\bibinfo {author} {\bibfnamefont {P.}~\bibnamefont
  {Corfdir}}, \bibinfo {author} {\bibfnamefont {H.}~\bibnamefont
  {K{\"{u}}pers}}, \bibinfo {author} {\bibfnamefont {R.~B.}\ \bibnamefont
  {Lewis}}, \bibinfo {author} {\bibfnamefont {T.}~\bibnamefont {Flissikowski}},
  \bibinfo {author} {\bibfnamefont {H.~T.}\ \bibnamefont {Grahn}}, \bibinfo
  {author} {\bibfnamefont {L.}~\bibnamefont {Geelhaar}}, \ and\ \bibinfo
  {author} {\bibfnamefont {O.}~\bibnamefont {Brandt}},\ }\href {\doibase
  10.1103/PhysRevB.94.155413} {\bibfield  {journal} {\bibinfo  {journal} {Phys.
  Rev. B}\ }\textbf {\bibinfo {volume} {94}},\ \bibinfo {pages} {155413}
  (\bibinfo {year} {2016}{\natexlab{b}})}\BibitemShut {NoStop}%
\bibitem [{\citenamefont {Rodt}\ \emph {et~al.}(2003)\citenamefont {Rodt},
  \citenamefont {Heitz}, \citenamefont {Schliwa}, \citenamefont {Sellin},
  \citenamefont {Guffarth},\ and\ \citenamefont {Bimberg}}]{Rodt2003}%
  \BibitemOpen
  \bibfield  {author} {\bibinfo {author} {\bibfnamefont {S.}~\bibnamefont
  {Rodt}}, \bibinfo {author} {\bibfnamefont {R.}~\bibnamefont {Heitz}},
  \bibinfo {author} {\bibfnamefont {A.}~\bibnamefont {Schliwa}}, \bibinfo
  {author} {\bibfnamefont {R.~L.}\ \bibnamefont {Sellin}}, \bibinfo {author}
  {\bibfnamefont {F.}~\bibnamefont {Guffarth}}, \ and\ \bibinfo {author}
  {\bibfnamefont {D.}~\bibnamefont {Bimberg}},\ }\href {\doibase
  10.1103/PhysRevB.68.035331} {\bibfield  {journal} {\bibinfo  {journal} {Phys.
  Rev. B}\ }\textbf {\bibinfo {volume} {68}},\ \bibinfo {pages} {035331}
  (\bibinfo {year} {2003})}\BibitemShut {NoStop}%
\bibitem [{\citenamefont {Uccelli}\ \emph {et~al.}(2008)\citenamefont
  {Uccelli}, \citenamefont {Bauer}, \citenamefont {Bichler}, \citenamefont
  {Schuh}, \citenamefont {Finley}, \citenamefont {Abstreiter},\ and\
  \citenamefont {{Fontcuberta i Morral}}}]{Uccelli2008}%
  \BibitemOpen
  \bibfield  {author} {\bibinfo {author} {\bibfnamefont {E.}~\bibnamefont
  {Uccelli}}, \bibinfo {author} {\bibfnamefont {J.}~\bibnamefont {Bauer}},
  \bibinfo {author} {\bibfnamefont {M.}~\bibnamefont {Bichler}}, \bibinfo
  {author} {\bibfnamefont {D.}~\bibnamefont {Schuh}}, \bibinfo {author}
  {\bibfnamefont {J.~J.}\ \bibnamefont {Finley}}, \bibinfo {author}
  {\bibfnamefont {G.}~\bibnamefont {Abstreiter}}, \ and\ \bibinfo {author}
  {\bibfnamefont {A.}~\bibnamefont {{Fontcuberta i Morral}}},\ }\enquote
  {\bibinfo {title} {{Self-assembly of InAs Quantum Dot Structures on Cleaved
  Facets}},}\ in\ \href {\doibase 10.1007/978-0-387-74191-8_2} {\emph {\bibinfo
  {booktitle} {{Self-Assembled Quantum Dots}}}},\ \bibinfo {editor} {edited by\
  \bibinfo {editor} {\bibfnamefont {Z.~M.}\ \bibnamefont {Wang}}}\ (\bibinfo
  {publisher} {Springer New York},\ \bibinfo {address} {New York, NY},\
  \bibinfo {year} {2008})\ pp.\ \bibinfo {pages} {25--41}\BibitemShut {NoStop}%
\bibitem [{\citenamefont {Uccelli}\ \emph {et~al.}(2010)\citenamefont
  {Uccelli}, \citenamefont {Arbiol}, \citenamefont {Morante},\ and\
  \citenamefont {{Fontcuberta i Morral}}}]{Uccelli2010}%
  \BibitemOpen
  \bibfield  {author} {\bibinfo {author} {\bibfnamefont {E.}~\bibnamefont
  {Uccelli}}, \bibinfo {author} {\bibfnamefont {J.}~\bibnamefont {Arbiol}},
  \bibinfo {author} {\bibfnamefont {J.~R.}\ \bibnamefont {Morante}}, \ and\
  \bibinfo {author} {\bibfnamefont {A.}~\bibnamefont {{Fontcuberta i
  Morral}}},\ }\href {\doibase 10.1021/nn101604k} {\bibfield  {journal}
  {\bibinfo  {journal} {ACS Nano}\ }\textbf {\bibinfo {volume} {4}},\ \bibinfo
  {pages} {5985} (\bibinfo {year} {2010})}\BibitemShut {NoStop}%
\bibitem [{\citenamefont {{Kanta Patra}}\ and\ \citenamefont
  {Schulz}(2017)}]{KantaPatra2017}%
  \BibitemOpen
  \bibfield  {author} {\bibinfo {author} {\bibfnamefont {S.}~\bibnamefont
  {{Kanta Patra}}}\ and\ \bibinfo {author} {\bibfnamefont {S.}~\bibnamefont
  {Schulz}},\ }\href {\doibase 10.1088/1361-6463/50/2/025108} {\bibfield
  {journal} {\bibinfo  {journal} {J. Phys. D: Appl. Phys.}\ }\textbf {\bibinfo
  {volume} {50}},\ \bibinfo {pages} {025108} (\bibinfo {year}
  {2017})}\BibitemShut {NoStop}%
\bibitem [{\citenamefont {Giddings}\ \emph {et~al.}(2011)\citenamefont
  {Giddings}, \citenamefont {Keizer}, \citenamefont {Hara}, \citenamefont
  {Hamhuis}, \citenamefont {Yuasa}, \citenamefont {Fukuzawa},\ and\
  \citenamefont {Koenraad}}]{Giddings2011}%
  \BibitemOpen
  \bibfield  {author} {\bibinfo {author} {\bibfnamefont {A.~D.}\ \bibnamefont
  {Giddings}}, \bibinfo {author} {\bibfnamefont {J.~G.}\ \bibnamefont
  {Keizer}}, \bibinfo {author} {\bibfnamefont {M.}~\bibnamefont {Hara}},
  \bibinfo {author} {\bibfnamefont {G.~J.}\ \bibnamefont {Hamhuis}}, \bibinfo
  {author} {\bibfnamefont {H.}~\bibnamefont {Yuasa}}, \bibinfo {author}
  {\bibfnamefont {H.}~\bibnamefont {Fukuzawa}}, \ and\ \bibinfo {author}
  {\bibfnamefont {P.~M.}\ \bibnamefont {Koenraad}},\ }\href {\doibase
  10.1103/PhysRevB.83.205308} {\bibfield  {journal} {\bibinfo  {journal} {Phys.
  Rev. B}\ }\textbf {\bibinfo {volume} {83}},\ \bibinfo {pages} {205308}
  (\bibinfo {year} {2011})}\BibitemShut {NoStop}%
\bibitem [{\citenamefont {Verheijen}\ \emph {et~al.}(2007)\citenamefont
  {Verheijen}, \citenamefont {Algra}, \citenamefont {Borgström}, \citenamefont
  {Immink}, \citenamefont {Sourty}, \citenamefont {van Enckevort},
  \citenamefont {Vlieg},\ and\ \citenamefont {Bakkers}}]{Verheijen2007}%
  \BibitemOpen
  \bibfield  {author} {\bibinfo {author} {\bibfnamefont {M.~A.}\ \bibnamefont
  {Verheijen}}, \bibinfo {author} {\bibfnamefont {R.~E.}\ \bibnamefont
  {Algra}}, \bibinfo {author} {\bibfnamefont {M.~T.}\ \bibnamefont
  {Borgström}}, \bibinfo {author} {\bibfnamefont {G.}~\bibnamefont {Immink}},
  \bibinfo {author} {\bibfnamefont {E.}~\bibnamefont {Sourty}}, \bibinfo
  {author} {\bibfnamefont {W.~J.~P.}\ \bibnamefont {van Enckevort}}, \bibinfo
  {author} {\bibfnamefont {E.}~\bibnamefont {Vlieg}}, \ and\ \bibinfo {author}
  {\bibfnamefont {E.~P. A.~M.}\ \bibnamefont {Bakkers}},\ }\href {\doibase
  10.1021/nl071541q} {\bibfield  {journal} {\bibinfo  {journal} {Nano Lett.}\
  }\textbf {\bibinfo {volume} {7}},\ \bibinfo {pages} {3051} (\bibinfo {year}
  {2007})}\BibitemShut {NoStop}%
\bibitem [{\citenamefont {Niehle}\ \emph {et~al.}(2015)\citenamefont {Niehle},
  \citenamefont {Trampert}, \citenamefont {Albert}, \citenamefont
  {Bengoechea-Encabo},\ and\ \citenamefont {Calleja}}]{Niehle2015}%
  \BibitemOpen
  \bibfield  {author} {\bibinfo {author} {\bibfnamefont {M.}~\bibnamefont
  {Niehle}}, \bibinfo {author} {\bibfnamefont {A.}~\bibnamefont {Trampert}},
  \bibinfo {author} {\bibfnamefont {S.}~\bibnamefont {Albert}}, \bibinfo
  {author} {\bibfnamefont {A.}~\bibnamefont {Bengoechea-Encabo}}, \ and\
  \bibinfo {author} {\bibfnamefont {E.}~\bibnamefont {Calleja}},\ }\href
  {\doibase 10.1063/1.4914102} {\bibfield  {journal} {\bibinfo  {journal} {APL
  Mater.}\ }\textbf {\bibinfo {volume} {3}},\ \bibinfo {pages} {036102}
  (\bibinfo {year} {2015})}\BibitemShut {NoStop}%
\bibitem [{\citenamefont {Walck}\ and\ \citenamefont
  {Reinecke}(1998)}]{Walck1998}%
  \BibitemOpen
  \bibfield  {author} {\bibinfo {author} {\bibfnamefont {S.~N.}\ \bibnamefont
  {Walck}}\ and\ \bibinfo {author} {\bibfnamefont {T.~L.}\ \bibnamefont
  {Reinecke}},\ }\href {\doibase 10.1103/PhysRevB.57.9088} {\bibfield
  {journal} {\bibinfo  {journal} {Phys. Rev. B}\ }\textbf {\bibinfo {volume}
  {57}},\ \bibinfo {pages} {9088} (\bibinfo {year} {1998})}\BibitemShut
  {NoStop}%
\bibitem [{\citenamefont {Ishii}\ \emph {et~al.}(2003)\citenamefont {Ishii},
  \citenamefont {Aisaka}, \citenamefont {Oh}, \citenamefont {Yoshita},\ and\
  \citenamefont {Akiyama}}]{Ishii2003}%
  \BibitemOpen
  \bibfield  {author} {\bibinfo {author} {\bibfnamefont {A.}~\bibnamefont
  {Ishii}}, \bibinfo {author} {\bibfnamefont {T.}~\bibnamefont {Aisaka}},
  \bibinfo {author} {\bibfnamefont {J.-W.}\ \bibnamefont {Oh}}, \bibinfo
  {author} {\bibfnamefont {M.}~\bibnamefont {Yoshita}}, \ and\ \bibinfo
  {author} {\bibfnamefont {H.}~\bibnamefont {Akiyama}},\ }\href {\doibase
  10.1063/1.1627945} {\bibfield  {journal} {\bibinfo  {journal} {Appl. Phys.
  Lett.}\ }\textbf {\bibinfo {volume} {83}},\ \bibinfo {pages} {4187} (\bibinfo
  {year} {2003})}\BibitemShut {NoStop}%
\bibitem [{\citenamefont {Pryor}\ and\ \citenamefont
  {Flatt\'e}(2006)}]{Pryor2006}%
  \BibitemOpen
  \bibfield  {author} {\bibinfo {author} {\bibfnamefont {C.~E.}\ \bibnamefont
  {Pryor}}\ and\ \bibinfo {author} {\bibfnamefont {M.~E.}\ \bibnamefont
  {Flatt\'e}},\ }\href {\doibase 10.1103/PhysRevLett.96.026804} {\bibfield
  {journal} {\bibinfo  {journal} {Phys. Rev. Lett.}\ }\textbf {\bibinfo
  {volume} {96}},\ \bibinfo {pages} {026804} (\bibinfo {year}
  {2006})}\BibitemShut {NoStop}%
\bibitem [{\citenamefont {Oberli}\ \emph {et~al.}(2009)\citenamefont {Oberli},
  \citenamefont {Byszewski}, \citenamefont {Chalupar}, \citenamefont
  {Pelucchi}, \citenamefont {Rudra},\ and\ \citenamefont {Kapon}}]{Oberli2009}%
  \BibitemOpen
  \bibfield  {author} {\bibinfo {author} {\bibfnamefont {D.~Y.}\ \bibnamefont
  {Oberli}}, \bibinfo {author} {\bibfnamefont {M.}~\bibnamefont {Byszewski}},
  \bibinfo {author} {\bibfnamefont {B.}~\bibnamefont {Chalupar}}, \bibinfo
  {author} {\bibfnamefont {E.}~\bibnamefont {Pelucchi}}, \bibinfo {author}
  {\bibfnamefont {A.}~\bibnamefont {Rudra}}, \ and\ \bibinfo {author}
  {\bibfnamefont {E.}~\bibnamefont {Kapon}},\ }\href {\doibase
  10.1103/PhysRevB.80.165312} {\bibfield  {journal} {\bibinfo  {journal} {Phys.
  Rev. B}\ }\textbf {\bibinfo {volume} {80}},\ \bibinfo {pages} {165312}
  (\bibinfo {year} {2009})}\BibitemShut {NoStop}%
\bibitem [{\citenamefont {Schwan}\ \emph {et~al.}(2011)\citenamefont {Schwan},
  \citenamefont {Meiners}, \citenamefont {Greilich}, \citenamefont {Yakovlev},
  \citenamefont {Bayer}, \citenamefont {Maia}, \citenamefont {Quivy},\ and\
  \citenamefont {Henriques}}]{Schwan2011}%
  \BibitemOpen
  \bibfield  {author} {\bibinfo {author} {\bibfnamefont {A.}~\bibnamefont
  {Schwan}}, \bibinfo {author} {\bibfnamefont {B.-M.}\ \bibnamefont {Meiners}},
  \bibinfo {author} {\bibfnamefont {A.}~\bibnamefont {Greilich}}, \bibinfo
  {author} {\bibfnamefont {D.~R.}\ \bibnamefont {Yakovlev}}, \bibinfo {author}
  {\bibfnamefont {M.}~\bibnamefont {Bayer}}, \bibinfo {author} {\bibfnamefont
  {A.~D.~B.}\ \bibnamefont {Maia}}, \bibinfo {author} {\bibfnamefont {A.~A.}\
  \bibnamefont {Quivy}}, \ and\ \bibinfo {author} {\bibfnamefont {A.~B.}\
  \bibnamefont {Henriques}},\ }\href {\doibase 10.1063/1.3665634} {\bibfield
  {journal} {\bibinfo  {journal} {Appl. Phys. Lett.}\ }\textbf {\bibinfo
  {volume} {99}},\ \bibinfo {pages} {221914} (\bibinfo {year}
  {2011})}\BibitemShut {NoStop}%
\bibitem [{\citenamefont {Van~Hattem}\ \emph {et~al.}(2013)\citenamefont
  {Van~Hattem}, \citenamefont {Corfdir}, \citenamefont {Brereton},
  \citenamefont {Pearce}, \citenamefont {Graham}, \citenamefont {Stanley},
  \citenamefont {Hugues}, \citenamefont {Hopkinson},\ and\ \citenamefont
  {Phillips}}]{VanHattem2013}%
  \BibitemOpen
  \bibfield  {author} {\bibinfo {author} {\bibfnamefont {B.}~\bibnamefont
  {Van~Hattem}}, \bibinfo {author} {\bibfnamefont {P.}~\bibnamefont {Corfdir}},
  \bibinfo {author} {\bibfnamefont {P.}~\bibnamefont {Brereton}}, \bibinfo
  {author} {\bibfnamefont {P.}~\bibnamefont {Pearce}}, \bibinfo {author}
  {\bibfnamefont {A.~M.}\ \bibnamefont {Graham}}, \bibinfo {author}
  {\bibfnamefont {M.~J.}\ \bibnamefont {Stanley}}, \bibinfo {author}
  {\bibfnamefont {M.}~\bibnamefont {Hugues}}, \bibinfo {author} {\bibfnamefont
  {M.}~\bibnamefont {Hopkinson}}, \ and\ \bibinfo {author} {\bibfnamefont
  {R.~T.}\ \bibnamefont {Phillips}},\ }\href {\doibase
  10.1103/PhysRevB.87.205308} {\bibfield  {journal} {\bibinfo  {journal} {Phys.
  Rev. B}\ }\textbf {\bibinfo {volume} {87}},\ \bibinfo {pages} {205308}
  (\bibinfo {year} {2013})}\BibitemShut {NoStop}%
\bibitem [{\citenamefont {Nakaoka}\ \emph {et~al.}(2005)\citenamefont
  {Nakaoka}, \citenamefont {Saito}, \citenamefont {Tatebayashi}, \citenamefont
  {Hirose}, \citenamefont {Usuki}, \citenamefont {Yokoyama},\ and\
  \citenamefont {Arakawa}}]{Nakaoka2005}%
  \BibitemOpen
  \bibfield  {author} {\bibinfo {author} {\bibfnamefont {T.}~\bibnamefont
  {Nakaoka}}, \bibinfo {author} {\bibfnamefont {T.}~\bibnamefont {Saito}},
  \bibinfo {author} {\bibfnamefont {J.}~\bibnamefont {Tatebayashi}}, \bibinfo
  {author} {\bibfnamefont {S.}~\bibnamefont {Hirose}}, \bibinfo {author}
  {\bibfnamefont {T.}~\bibnamefont {Usuki}}, \bibinfo {author} {\bibfnamefont
  {N.}~\bibnamefont {Yokoyama}}, \ and\ \bibinfo {author} {\bibfnamefont
  {Y.}~\bibnamefont {Arakawa}},\ }\href {\doibase 10.1103/PhysRevB.71.205301}
  {\bibfield  {journal} {\bibinfo  {journal} {Phys. Rev. B}\ }\textbf {\bibinfo
  {volume} {71}},\ \bibinfo {pages} {205301} (\bibinfo {year}
  {2005})}\BibitemShut {NoStop}%
\bibitem [{\citenamefont {Simeonov}\ \emph {et~al.}(2008)\citenamefont
  {Simeonov}, \citenamefont {Dussaigne}, \citenamefont {Butt\'e},\ and\
  \citenamefont {Grandjean}}]{Simeonov2008}%
  \BibitemOpen
  \bibfield  {author} {\bibinfo {author} {\bibfnamefont {D.}~\bibnamefont
  {Simeonov}}, \bibinfo {author} {\bibfnamefont {A.}~\bibnamefont {Dussaigne}},
  \bibinfo {author} {\bibfnamefont {R.}~\bibnamefont {Butt\'e}}, \ and\
  \bibinfo {author} {\bibfnamefont {N.}~\bibnamefont {Grandjean}},\ }\href
  {\doibase 10.1103/PhysRevB.77.075306} {\bibfield  {journal} {\bibinfo
  {journal} {Phys. Rev. B}\ }\textbf {\bibinfo {volume} {77}},\ \bibinfo
  {pages} {075306} (\bibinfo {year} {2008})}\BibitemShut {NoStop}%
\bibitem [{\citenamefont {Salehzadeh}\ \emph {et~al.}(2013)\citenamefont
  {Salehzadeh}, \citenamefont {Kavanagh},\ and\ \citenamefont
  {Watkins}}]{Salehzadeh2013}%
  \BibitemOpen
  \bibfield  {author} {\bibinfo {author} {\bibfnamefont {O.}~\bibnamefont
  {Salehzadeh}}, \bibinfo {author} {\bibfnamefont {K.~L.}\ \bibnamefont
  {Kavanagh}}, \ and\ \bibinfo {author} {\bibfnamefont {S.~P.}\ \bibnamefont
  {Watkins}},\ }\href {\doibase 10.1063/1.4816460} {\bibfield  {journal}
  {\bibinfo  {journal} {J. Appl. Phys.}\ }\textbf {\bibinfo {volume} {114}},\
  \bibinfo {pages} {054301} (\bibinfo {year} {2013})}\BibitemShut {NoStop}%
\bibitem [{\citenamefont {Avron}\ \emph {et~al.}(2008)\citenamefont {Avron},
  \citenamefont {Bisker}, \citenamefont {Gershoni}, \citenamefont {Lindner},
  \citenamefont {Meirom},\ and\ \citenamefont {Warburton}}]{Avron2008}%
  \BibitemOpen
  \bibfield  {author} {\bibinfo {author} {\bibfnamefont {J.~E.}\ \bibnamefont
  {Avron}}, \bibinfo {author} {\bibfnamefont {G.}~\bibnamefont {Bisker}},
  \bibinfo {author} {\bibfnamefont {D.}~\bibnamefont {Gershoni}}, \bibinfo
  {author} {\bibfnamefont {N.~H.}\ \bibnamefont {Lindner}}, \bibinfo {author}
  {\bibfnamefont {E.~A.}\ \bibnamefont {Meirom}}, \ and\ \bibinfo {author}
  {\bibfnamefont {R.~J.}\ \bibnamefont {Warburton}},\ }\href {\doibase
  10.1103/PhysRevLett.100.120501} {\bibfield  {journal} {\bibinfo  {journal}
  {Phys. Rev. Lett.}\ }\textbf {\bibinfo {volume} {100}},\ \bibinfo {pages}
  {120501} (\bibinfo {year} {2008})}\BibitemShut {NoStop}%
\bibitem [{\citenamefont {H{\"{o}}nig}\ \emph {et~al.}(2014)\citenamefont
  {H{\"{o}}nig}, \citenamefont {Callsen}, \citenamefont {Schliwa},
  \citenamefont {Kalinowski}, \citenamefont {Kindel}, \citenamefont {Kako},
  \citenamefont {Arakawa}, \citenamefont {Bimberg},\ and\ \citenamefont
  {Hoffmann}}]{Honig2014}%
  \BibitemOpen
  \bibfield  {author} {\bibinfo {author} {\bibfnamefont {G.}~\bibnamefont
  {H{\"{o}}nig}}, \bibinfo {author} {\bibfnamefont {G.}~\bibnamefont
  {Callsen}}, \bibinfo {author} {\bibfnamefont {A.}~\bibnamefont {Schliwa}},
  \bibinfo {author} {\bibfnamefont {S.}~\bibnamefont {Kalinowski}}, \bibinfo
  {author} {\bibfnamefont {C.}~\bibnamefont {Kindel}}, \bibinfo {author}
  {\bibfnamefont {S.}~\bibnamefont {Kako}}, \bibinfo {author} {\bibfnamefont
  {Y.}~\bibnamefont {Arakawa}}, \bibinfo {author} {\bibfnamefont
  {D.}~\bibnamefont {Bimberg}}, \ and\ \bibinfo {author} {\bibfnamefont
  {A.}~\bibnamefont {Hoffmann}},\ }\href {\doibase 10.1038/ncomms6721}
  {\bibfield  {journal} {\bibinfo  {journal} {Nat. Commun.}\ }\textbf {\bibinfo
  {volume} {5}},\ \bibinfo {pages} {5721} (\bibinfo {year} {2014})}\BibitemShut
  {NoStop}%
\bibitem [{\citenamefont {Kajikawa}\ \emph {et~al.}(1992)\citenamefont
  {Kajikawa}, \citenamefont {Hata}, \citenamefont {Isu},\ and\ \citenamefont
  {Katayama}}]{Kajikawa1991}%
  \BibitemOpen
  \bibfield  {author} {\bibinfo {author} {\bibfnamefont {Y.}~\bibnamefont
  {Kajikawa}}, \bibinfo {author} {\bibfnamefont {M.}~\bibnamefont {Hata}},
  \bibinfo {author} {\bibfnamefont {T.}~\bibnamefont {Isu}}, \ and\ \bibinfo
  {author} {\bibfnamefont {Y.}~\bibnamefont {Katayama}},\ }\href {\doibase
  http://dx.doi.org/10.1016/0039-6028(92)91186-F} {\bibfield  {journal}
  {\bibinfo  {journal} {Surf. Sci.}\ }\textbf {\bibinfo {volume} {267}},\
  \bibinfo {pages} {501 } (\bibinfo {year} {1992})}\BibitemShut {NoStop}%
\bibitem [{\citenamefont {Muller}\ \emph {et~al.}(1993)\citenamefont {Muller},
  \citenamefont {Breguet},\ and\ \citenamefont {Gisin}}]{Muller1993}%
  \BibitemOpen
  \bibfield  {author} {\bibinfo {author} {\bibfnamefont {A.}~\bibnamefont
  {Muller}}, \bibinfo {author} {\bibfnamefont {J.}~\bibnamefont {Breguet}}, \
  and\ \bibinfo {author} {\bibfnamefont {N.}~\bibnamefont {Gisin}},\ }\href
  {http://stacks.iop.org/0295-5075/23/i=6/a=001} {\bibfield  {journal}
  {\bibinfo  {journal} {Europhys. Lett.}\ }\textbf {\bibinfo {volume} {23}},\
  \bibinfo {pages} {383} (\bibinfo {year} {1993})}\BibitemShut {NoStop}%
\end{thebibliography}%

\end{document}